# A memristive model for graphene emitters: hysteresis and self-crossing

**D. V. Gorodetskiy[a,*], S. N. Shevchenko[b,c], A. V. Gusel'nikov[a], A. V. Okotrub[a]**
[a]Nikolaev Institute of Inorganic Chemistry SB RAS, Novosibirsk 630090, Russia
[b]B. Verkin Institute for Low Temperature Physics and Engineering, Kharkov 61103, Ukraine
[c]V. N. Karazin Kharkov National University, Kharkov 61022, Ukraine

**Abstract.** Hysteresis exhibited by current-voltage characteristics during field-emission experiments is often considered undesirable in terms of practical applications. However, this is an appealing effect for the purposes of memristive devices. We developed a two-stage model to describe hysteretic characteristics, with a particular focus on the system which includes a cathode made of a single-layered graphene sheet on a substrate. In addition to hysteresis, the current-voltage curves display also an unusual self-crossing behavior. The presented memristive model can be used for quantitative descriptions of different hysteretic characteristics such as abrupt changes and self-crossings and for understanding (and modeling) the processes associated with field emission from plane graphene emitters.

* D. V. Gorodetskiy, gorodetskiy@niic.nsc.ru

## 1. Introduction

Graphene is an interesting material that combines a variety of physical and chemical properties. Its application areas are sometimes quite surprising. One possible direction of graphene research is field emission occurring at the edges of graphene sheets. Here, too, graphene shows some interesting features. Besides high aspect ratio, field emission from graphene cathodes demonstrates another intriguing feature, namely, huge hysteresis in current-voltage dependences.[1-5] Hysteresis in current-voltage characteristics of field emission may be due to a variety factors such as the emitter temperature, sorption of residual gases on the emitter's surface, and mechanical changes of the emitter's shape under the influence of an electromagnetic field.[6-10]

Hysteresis in field emission is ubiquitous (e.g. see References [4, 11-14]). It is therefore important to develop a transparent and functional theory that would describe different hysteretic dependences possibly exhibiting nontrivial behavior such as abrupt changes or self-crossings. In this work, experimental data were used to develop a mathematical model that qualitatively describes field emission from graphene sheets. The model describes well the hysteresis in the I–V characteristic characterized by self-intersection and can be used to understand the processes accompanying electron emission by planar emitters having physical and chemical properties similar to those of graphene.

Hysteresis in current-voltage characteristics is useful for memory devices commonly referred to as memristors.[11, 15-16] Memristive devices are novel elements of electric circuits, in which the relation between output and input signals is determined by a dynamic internal state variable.[16] For definiteness, suppose there is an element with voltage $U(t)$ (the input), current $I(t)$ (the output), and an internal variable $x(t)$ (to be discussed later). Then the memory element, the memristor, is defined by the following general relations

$$I(t) = G_M(x, U, t)U(t), \qquad (1)$$

$$\dot{x} = f(x, U, t), \qquad (2)$$

where $G_M$ is the conductance (also called memductance, from "memory conductance"), which depends both on the input voltage and on the internal system state; this function is the inverse of memory resistance, the memristance. The dynamics of the internal variable $x$ is defined by function $f$.

It will be shown on the example of field-emission from a graphene cathode that experimental results can be conveniently described by our model. In particular, both our own experimental results and those reported recently in Reference[4] will be discussed. It is important to note that this approach can be adapted also to other systems, e.g. nanomechanical systems like those discussed in References[17-20] (see also Reference[21] and discussion in Reference[16]).

Thus, measuring current-voltage characteristics of graphene experimentally and developing a theoretical model that would be well comparable with experimental data is an important problem. In this work, we present a comprehensive study of hysteretic field emission from graphene cathodes; theoretical, technological, and experimental aspects are discussed.

## 2. Experimental realization of graphene field emitter

Field emission characteristics were measured using graphene by Graphenea, CIC nanoGUNE, Spain. The graphene was synthesized by chemical vapor deposition (CVD) on a copper substrate and then transferred to a silicon substrate. A 10×10 mm area of the silicon substrate was covered with a natural $SiO_2$ layer.

The sample was characterized by scanning electron microscopy (SEM) on a dual-beam FIB/FEI Helios 450S system and by Raman spectroscopy on a HORIBO LabRAM HR Evolution spectrometer with an Ar+ laser excitation of 514 nm. The SEM image in **Figure 1(a)** shows the graphene surface structure. The micrographs were taken at an accelerating voltage of ~ 1 kV. As can bee seen, there are minor areas of two-three layer graphene, which occupy less than 3% of the sample surface.

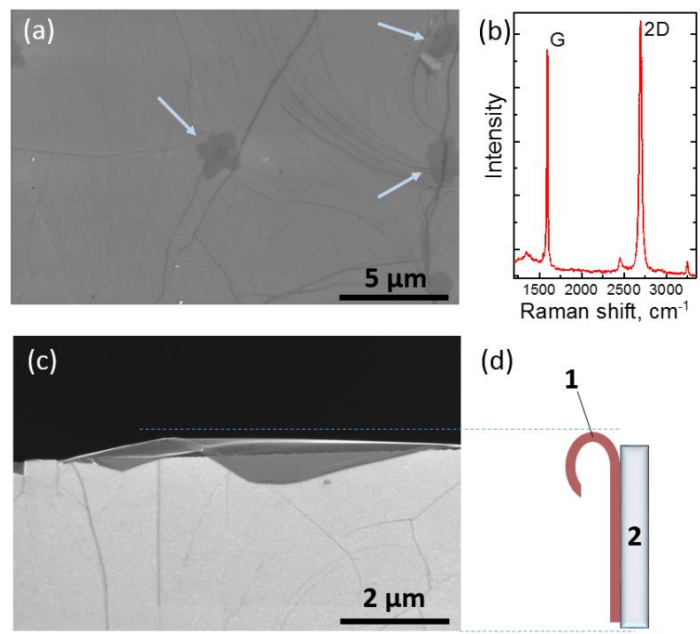

**Figure 1:** SEM image of the surface of a silicon substrate coated by graphene, the arrows show the regions of by graphene (a); Raman spectra of graphene (b); SEM image of the edge of the silicon substrate coated with graphene, top view (c); the same shown schematically; graphene and substrate designated by digits 1 and 2, respectively, side view (d).

**Figure 1(b)** shows the graphene Raman spectrum. The laser spot diameter was about 20 μm, the radiation power was 1 mW. The spectrum demonstrates G and 2D bands typical of graphene. The integrated intensities of G and 2D bands have approximately equal values to confirm a presence of several well-graphitized layers of graphene on the substrate surface.

In order to obtain smooth graphene edges, the substrates were mechanically cut into rectangular pieces (**Figure 1(c, d)**). **Figure 1(c)** shows the SEM image of the edges of the sample after splitting. As can be seen, graphene was broken irregularly and partially protruded beyond the silicon substrate. The protruding graphene edge has a size of about 6×2 µm. The radius of curvature of the ring-shaped graphene edge is about 0.5 µm. The side view of the sample is schematically shown in **Figure 1(d)**.

## 3. Current-voltage characteristics

The current-voltage characteristics were measured on our home-made set-up for field emission probing of carbon nanomaterials.[22] A half of the silicon substrate was placed in the $10^{-5}$ Torr vacuum chamber of the installation so that its fresh cut was directed towards the anode. The cut edge of graphene was parallel to the anode surface. The measurement scheme is shown in **Figure 2(a)**. An excitation voltage of 0—2.5 kV in diode configuration was applied to the sample at a frequency of 5 Hz. The sensitivity of the measuring current of the field emission was of the order of nA. The distance between the sample and the anode was controlled by a micromechanical screw system and was set equal to 1 mm.

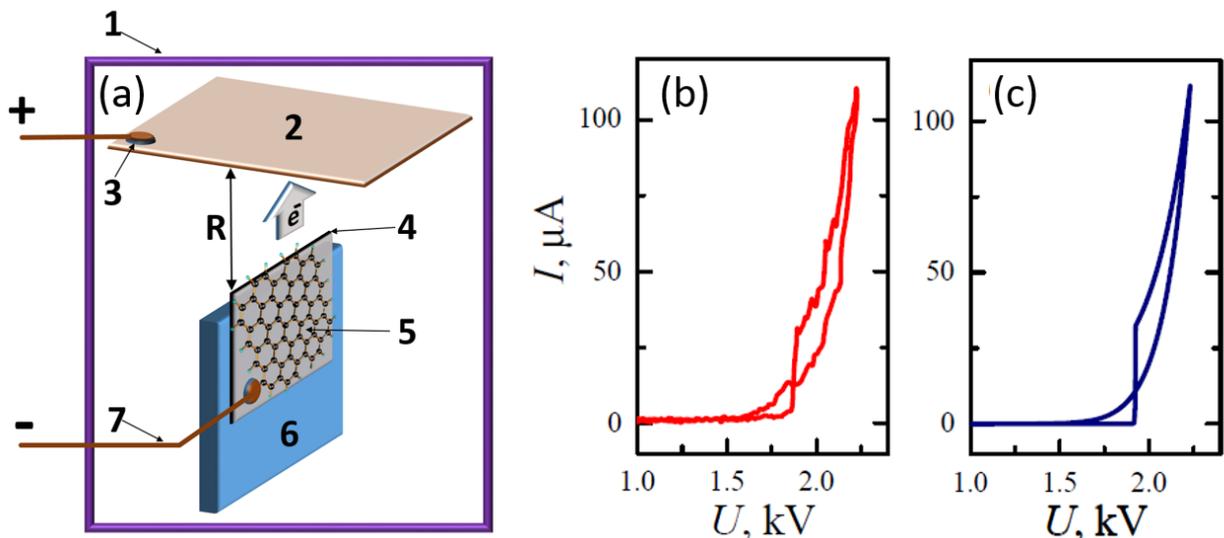

**Figure 2:** (a) Schematic experimental setup for field emission measurements: vacuum chamber of the field emission measurement system (1), flat anode (2), positive potential contact (3), emitting edge of graphene (4), silicon substrate coated with graphene (5), mobile substrate holder (6), negative electrode contact (7). (b, c) Hysteretic self-crossing of current-voltage characteristics: experiment (b) and theory (c). The rapid change of the form factor and the self-crossing point are shown.

**Figure 2(b)** shows I–V characteristics of several consecutive cycles. Note that the hysteretic curve has a self-intersection point. **Figure 4(c)** shows a calculated I–V characteristic with a self-intersecting hysteretic curve to be discussed in the next Section.

## 4. Modeling the self-crossing hysteresis

### 4.1. Two-stage memristive model

Let us describe the dynamics of the internal variable $x(t)$. First, we assume that this value varies from 0 to 1. We also require that it switches from 0 to 1 at $U = U_A$ as the voltage increases and back from 1 to 0 at $U = U_B < U_A$ as the voltage decreases. Ler $r$ be the rate of this transition, then the corresponding transition time is equal to $r^{-1}$. These requirements fulfilled, we obtain a dependence (**Figure 3**) that can be written as equation

$$\dot{x} = \vartheta\{\theta(U - U_A)\theta(1 - x) - \theta(U_B - U)\theta(x)\} \equiv f(x, U, t), \quad (3)$$

which corresponds to the second memristor relation, Equation 2. According to Reference[4], we take $U_A = 0.35$ kV, $U_B = 0.05$ kV and assume that the voltage varies periodically between 0 and $U_{max} = 0.5$ kV.

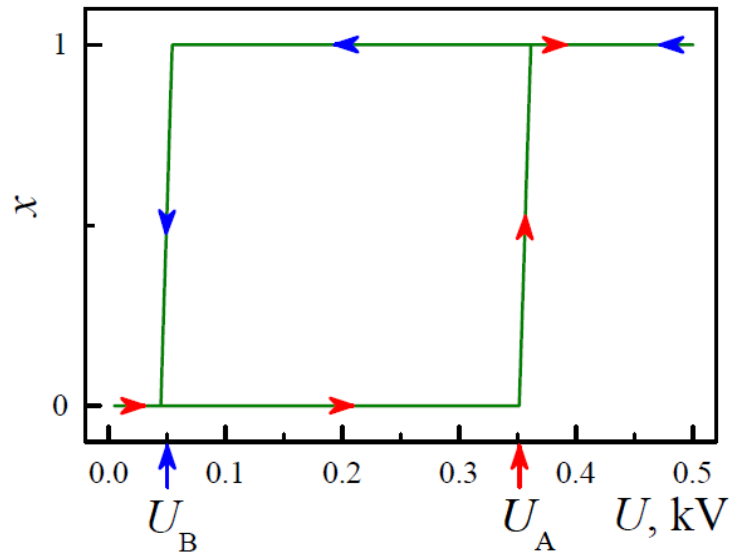

**Figure 3**: Two-stage model with a memristor internal variable $x$ rapidly switching between two input values. For definiteness, the changes in this system are assumed to occur at $U = U_{A;B}$.

## 4.2. Hysteretic current-voltage characteristics

To start with, we summarize some main features; more details can be found in the comprehensive analysis of Reference [4]. (i) The current-voltage characteristics for the blade-type graphene emitter obey the Fowler-Nordheim law (see also References[23-25]). (ii) Strong hysteresis cannot be explained neither by absorption of residual gas atoms and molecules nor by heating. (iii) Instead, it is explained by mechanical peeling of graphene sheet from the substrate. (iv) The effective emission surface (proportional to the number of emitting sites) changes for the increasing-voltage curve and remains virtually the same for the decreasing-voltage curve. (v) The self-crossing of the hysteretic curve can be observed at high voltages. Based on these observations, a model for quantitative description of such phenomena will be developed in this Section.

Let a potential difference $U(t)$ be applied between the cathode and the anode. It causes electric field $E = \beta U/D$, where $D$ is the distance between the electrodes and $\beta$ is the field enhancement factor (form factor). Then the emission current density at relatively low temperatures $T \ll \varphi/k_B$ is described by the Fowler—Nordheim formula[26, 27]

$$I(U) = AU^2 \exp\left(-\frac{B}{U}\right), \tag{4}$$

$$A = \frac{e^3}{16\pi^2 \hbar} \frac{1}{\varphi} \left(\frac{\beta}{D}\right)^2 S, \quad B = \frac{4\sqrt{2m}}{3e\hbar} \varphi^{3/2} \left(\frac{\beta}{D}\right)^{-1}, \tag{5}$$

where $e$ is the electron charge, $m$ is the electron mass, $\hbar$ is Planck's constant, $k_B$ is the Boltzmann constant; the dimension of quantity $S$ is the same as that of area and can be treated in the first approximation as the area of the emitting surface; the work function is assumed here constant and equal to $\varphi = 4.8$ eV.

Thus, formula (4) contains two fitting parameters $S$ and $b = \beta/D$. It was assumed in the experiments that the former is either constant or a slowly changing parameter, while the latter changes rapidly at $U = U_{A,B}$ together with the form factor when the graphene sheet delaminates and sticks back. For simplicity, the changes of the emitting surface $S$ will be neglected and the variable $x$ is assumed to describe the variation of parameter $b$ (or form factor $\beta$) so that $x = 0$ (1) corresponds to low (high) form factor values and can be termed as high-resistance (low-resistance) states. As a result, we obtain a memristive model of field emission, which can be rewritten in the form of equations (1-2) with

$$G_M(x, U) = A(x)U\exp\left(-\frac{B(x)}{U}\right), \tag{6}$$

$$A(x) = A_0(1 - x) + A_1 x, \quad B(x) = B_0(1 - x) + B_1 x.$$

This means that the increase of the form factor (i.e. graphene sheet delamination) results in transitions from $B(0) = B_0$ to $B(1) = B_1 < B_0$ and from $A_0$ to $A_1 > A_0$.

Alternatively, delamination of a large-area graphene sheet can be described as a continuous process whereby some part of the sheet (with a relative weight $x$) is peeled off while its other part (with a relative weight $1 - x$) remains stuck. This would result in the total conductance

$$G_M = G_{M0} + G_{M1} \equiv (1 - x)A_0 U e^{-\frac{B_0}{U}} + x A_1 U e^{-\frac{B_1}{U}}. \tag{7}$$

According to our calculations, the dependences furnished by this model are very similar to those of the former model, Equation 6. Therefore, for definiteness, we will below explore the model of Equation 6. Note that in Reference[15] it was more convenient to use the model of Equation 7.

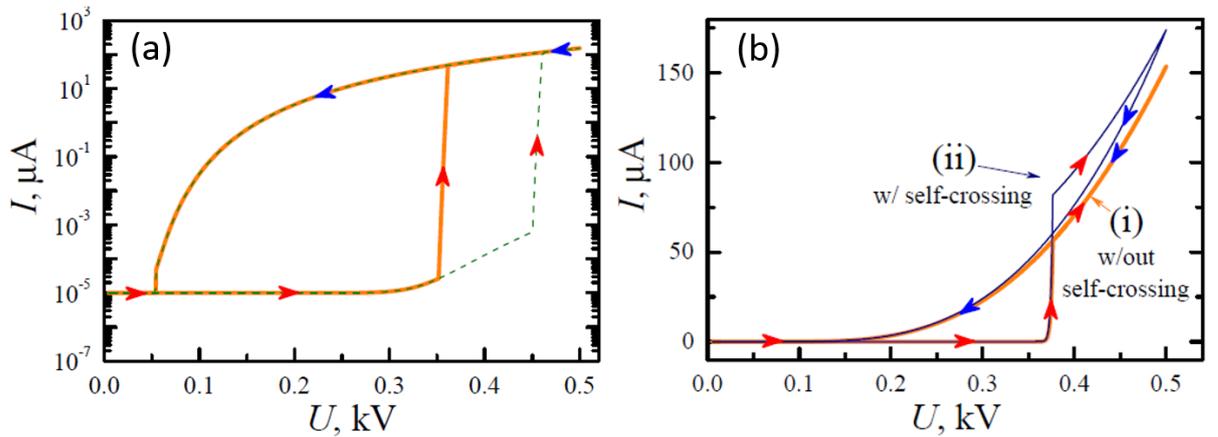

**Figure 4**: (a) Hysteretic current-voltage cycles with $U_B = 0.05$ and $U_A = 0.35$ and $0.45$ for thick solid and thin dashed lines, respectively. (b) Current-voltage characteristics in the linear scale (thick orange line (i)) as described by the rapidly changing parameter $x$. A self-crossing loop appearing as a result of non-equilibrium changes caused by the effect of inertia (thin blue line (ii)).

**Figure 4(a)** shows a strong hysteretic current-voltage cycle with the following parameters: $A_1 = 2300$, $B_1 = 0.66$, $A_0 = 200$, and $B_0 = 5$. We assume here that $I$ and $U$ are given in µA and kV, respectively, so that coefficients $A_i$ and $B_i$ are dimensionless. The switching voltage $U_A$ has two values so that these two curves are related to two experimental results presented in Reference[4] in Figures 3 and 5, respectively.

### 4.3. Self-crossing

The calculation results are presented in the logarithmic scale in **Figure 4(a)**. However, it is sometimes more convenient to use the linear scale (**Figure 4(b)**) for clarity reasons (cf. also the hysteresis in Reference[28]) so that the high-voltage region becomes better visible. Note that the self-crossing of hysteretic curves, which is barely visible in Figures 3 and 5 in Reference[4], is more pronounced in the linear scale.

To describe the self-crossing, we need to consider the effect of non-equilibrium. We consider here a visual model, which assumes, in addition to the previous consideration, that the peeled-off graphene sheet in the course of delamination moves at a slightly more distant position than that described by $x = 1$ an corresponds to some $x_M > 1$. This happens after the voltage reaches $U = U_A$. **Figure 5(a)** shows time dependence of parameter $x$ where time is normalized with respect to the driving frequency $\nu$ ($\tau = \nu\, t$). The second stage starts at $\tau_A$ and proceeds until $\tau_A + \delta$, where $\delta = \nu\, r^{-1}$. Then there is a relaxation stage that proceeds with rate $\Gamma$ and is described by $\gamma = \Gamma/\nu$. Finally, $x$ rapidly drops from 1 to 0 at $\tau = \tau_B$ (when $U = U_B$) to describe graphene sheet sticking back to the substrate. We believe that this model provides clear physical interpretation of most relevant and non-trivial features of hysteretic curves; we avoided overcomplicating in other possible changes during the variation of the voltage. Further, **Figure 5(b)** shows the dependence of parameter $x$ on the voltage, where the values $U_A$ and $U_B$ are defined in accordance with the experiment, as described below. By comparing **Figure 5(b)** with **Figure 3** one can explicitly see a self-crossing loop appearing in the model.

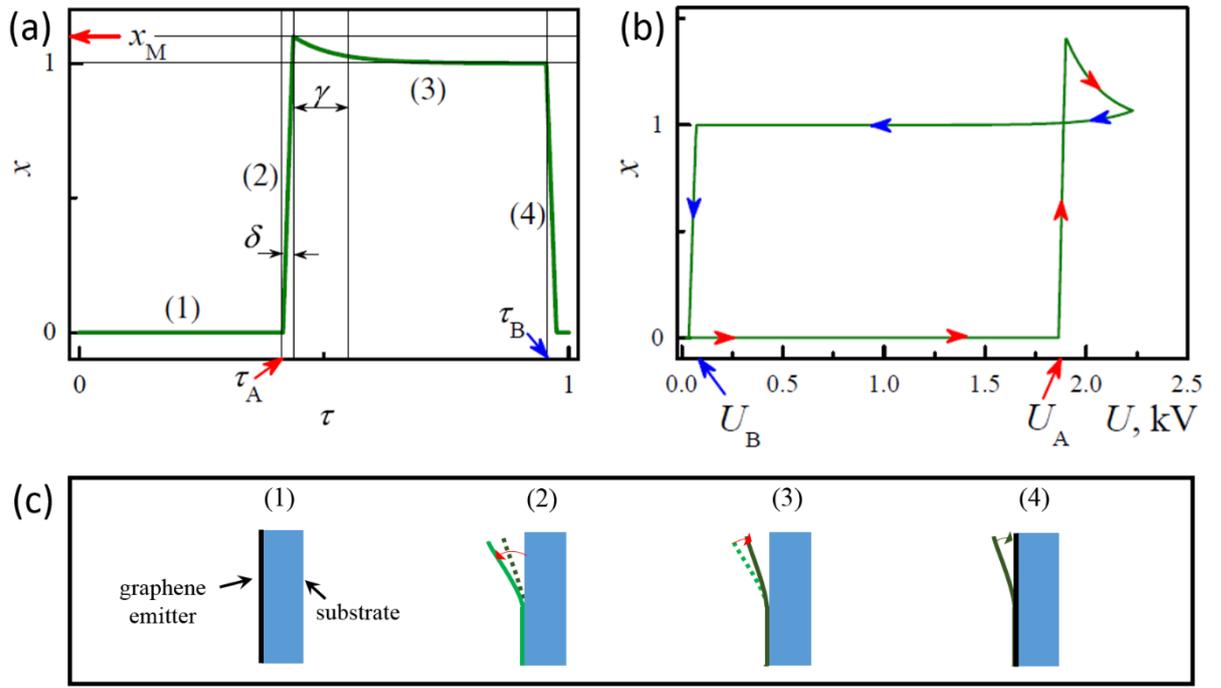

**Figure 5**: (a) Four-stage model for parameter $x$: $x = 0$ (graphene sheet is attached to the substrate) (1); $x$ rapidly switches from 0 to 1 and to a still higher value $x_M > 1$ due to inertia (graphene is peeled off from the substrate, the form factor is abruptly increased) (2); relaxation to 1 (graphene sheet returns to its equilibrium delaminated position) (3); $x$ rapidly switches from 1 to 0 (as the voltage decreases, the graphene sheet sticks back to the substrate) (4). (b) Voltage dependence of parameter $x$ in the model with a self-crossing point. (c) The above four stages shown schematically.

Let us now consider the model presented in Figure 2(c), having in mind to compare this with our in-home experiment. This theoretical graph was plotted according to Equation 6, the dependence $x = x(U(\tau))$ was taken from Figure 5. The following parameters were used: A1 = 1000, B1 = 25, A0 = A1/9, and B0 = 3B1 (corresponding to a three-fold increase of the form-factor, cf. Equation 5); UA = 1.86 kV, UB = 0.05 kV, xM = 1.05, r = 0.36 kHz, $\Gamma$ = 0.06 kHz. Similarly, **Figure 4(b)** shows a dotted line is not only changing abruptly but also contains a self-crossing point due to the non-equilibrium described by $x_M > 1$ (more specifically, $x_M$ was taken to be equal to 1.03).

## 5. Conclusion

A model describing hysteretic current-voltage cycles was developed for the field emission from a graphene cathode. The mechanical motion of graphene being peeled off from the substrate was described in terms of a two-stage memristive model that was shown useful when compared to realistic experiments. On the other hand, such layout may be useful for memristive applications. Having small operation frequencies and being reliably realized and controlled, they can be used to model their more practical counterparts, fast and small nano-scale memristors. Some possible applications to the realization of logic operations are discussed elsewhere.[15]

*Acknowledgments*

This work was supported by the Russian Science Foundation (grant No. 15-13-20021).